\documentclass[12pt]{article}
\usepackage{pic03}
\usepackage{hyperref}
\usepackage{url}
\usepackage{graphicx}

\begin{document}

\title{\bf STATUS OF THE H.E.S.S.\ EXPERIMENT}
\author{
Ullrich Schwanke\\
{\em Humboldt University Berlin, Department of Physics}\\
{\em Newtonstr. 15, D-12489 Berlin, Germany}\\
for the H.E.S.S.\ Collaboration\\
{\em http://www.mpi-hd.mpg.de/hfm/HESS/HESS.html}
}
\maketitle

\baselineskip=14.5pt
\begin{abstract}
The H.E.S.S.\ experiment (High Energy Stereoscopic System) is a new
generation atmospheric Cherenkov detector measuring high-energy ($>100$\,GeV)
gamma radiation from the Universe. This paper briefly explains the imaging 
Cherenkov technique and describes the H.E.S.S.\ project 
and its current status.

\end{abstract}

\baselineskip=17pt

\section{Introduction}

The detection of high-energy gamma rays in ground-based 
experiments opened a new window for the exploration of non-thermal
processes in the Universe. Unlike the charged component of the
cosmic radiation, detected gamma rays point back to their origin.
This property makes them an excellent tool for locating the 
sources of the cosmic radiation where hadron acceleration
is likely accompanied by gamma ray emission. The 
measured energy spectra of sources test our understanding 
of stellar objects and theoretical models of the acceleration 
mechanisms. Among others, active core regions of galaxies and remnants of supernovae
explosions are discussed as acceleration sites. The first 
source of TeV gamma radiation, the Crab nebula, was discovered in 1989.
To date, a handful of TeV gamma sources are known to exist.

\section{The Imaging Atmospheric Cherenkov Technique}

\begin{figure}[htbp]
  \centerline{\hbox{ \hspace{0.2cm}
    \includegraphics[width=0.47\linewidth]{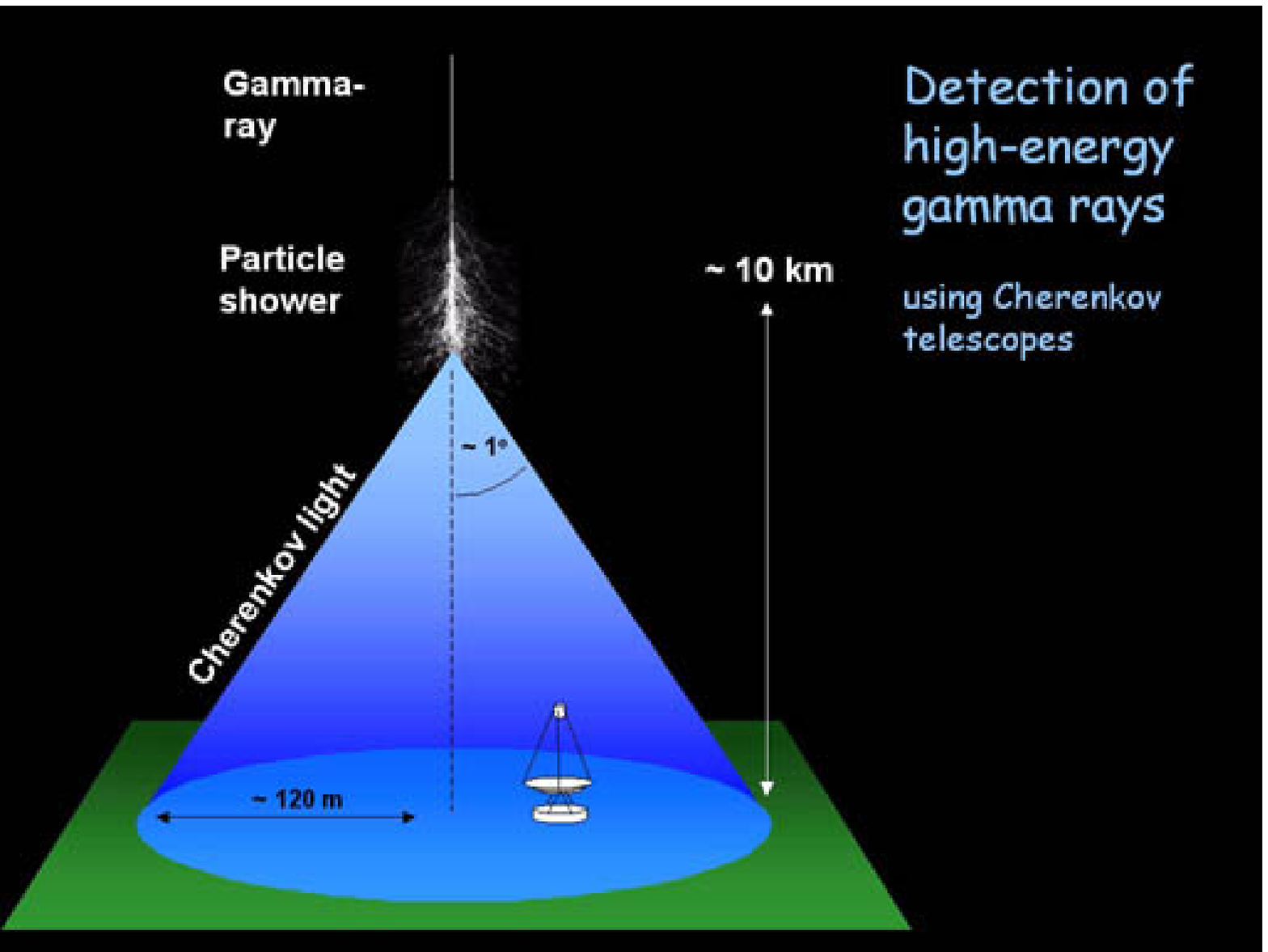}
    \hspace{0.3cm}
    \includegraphics[width=0.47\linewidth]{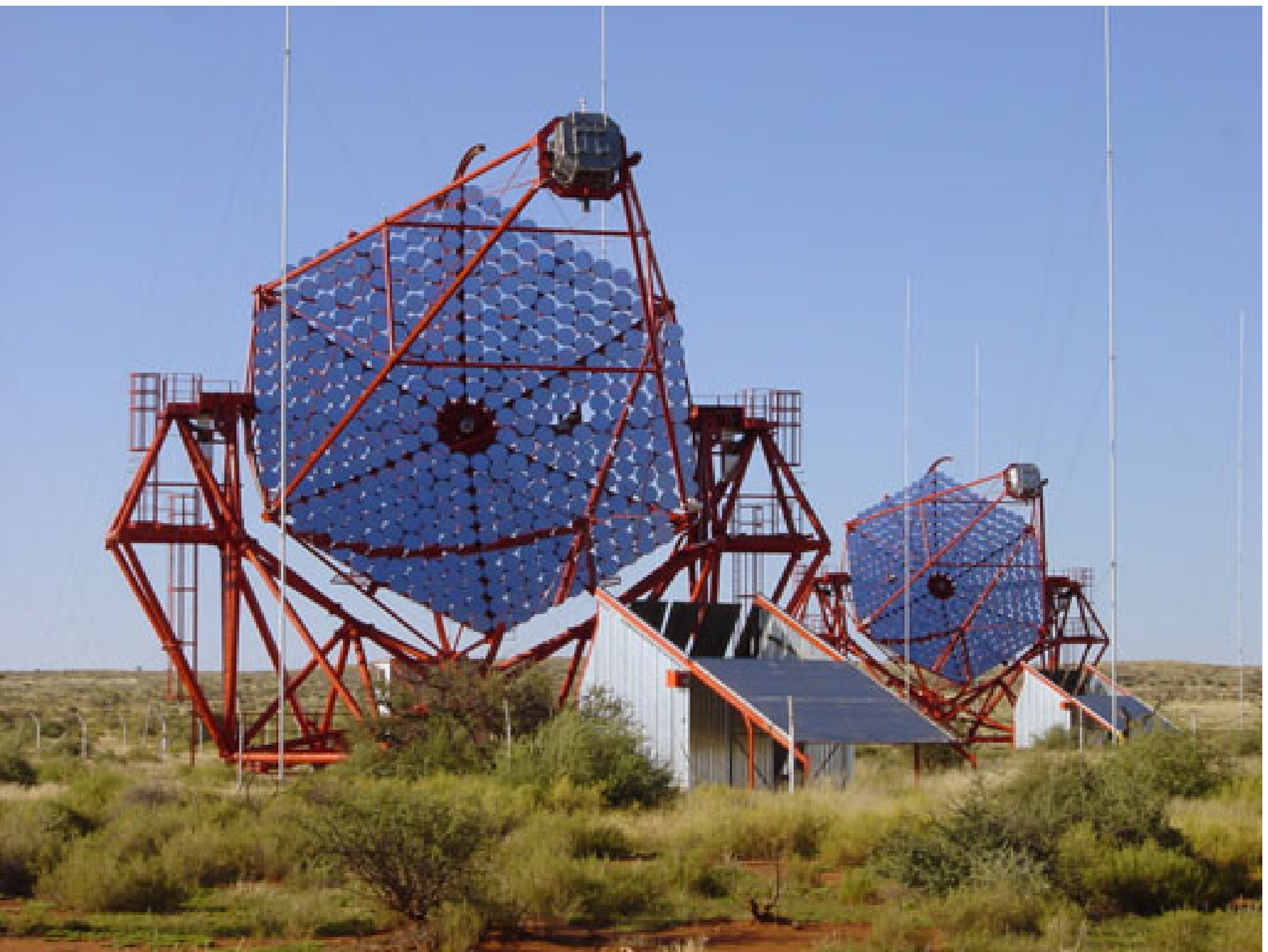}
    }
  }
 \caption{\it Illustration of the imaging atmospheric Cherenkov
    technique (left) and picture of two of the four Phase~I H.E.S.S.\ telescopes.
    See text for explanations.
    \label{fig:figure} }
\end{figure}

High-energy gamma rays and hadrons hitting the Earth's atmosphere 
create extended air showers. The shower development starts  
in a height of about 25\,km and the shower maximum
is reached at a height of around 8\,km. The charged secondary particles
in the shower produce Cherenkov light that forms a light pool 
with about 240\,m diameter on the ground (Fig.~\ref{fig:figure} (left)).
This light can be detected in clear moonless nights by 
so-called Imaging Atmospheric Cherenkov Telescopes (IACTs) placed in the light pool. 
The telescopes use a spherical mirror to map the shower image onto a 
digital camera. Analysis of the camera image allows to infer the
type, energy and direction of the primary particle.

\section{The H.E.S.S.\ Experiment}

The H.E.S.S.\ experiment \cite{hess1, hess2} applies the so-called stereoscopic observation mode 
where the same shower is recorded under different angles by more than one 
telescope in the light pool. This technique allows for better 
reconstruction of shower directions and improved separation of 
gamma-induced showers from the more irregular hadronic showers.

The H.E.S.S.\ experiment is located at 1800\,m above sea level 
in the Khomas Highland of Namibia
which was chosen for its good optical conditions and favourable 
climate. In its first phase, the experiment will consist of four IACTs located on the corners
of a 120\,m square. Each telescope (Fig.~\ref{fig:figure} (right)) 
is movable in azimuth and altitude and has a focal length of 15\,m. The
spherical mirror (diameter 15\,m) is made up of 380 individual mirrors
with a diameter of 60\,cm. The total mirror area is 107\,m$^2$ per
telescope. The camera has a diameter of 1.5\,m and a 5$^\circ$ 
field of view. It consists of 960 photo-multiplier tubes each of 
which covering 0.16$^\circ$. The camera houses all the required trigger
and readout electronics.

The sensitivity of the H.E.S.S.\ experiment is expected to be an order 
of magnitude better than that of existing IACT systems. Sources
with 1\,\% of the flux of the Crab nebula will be detectable. The energy 
threshold is about 100\,GeV and the energies of primary particles can
be reconstructed with an accuracy of 20\,\%. The resolution of 
shower directions is 0.1$^\circ$ per event. 

\section{Status}

The first H.E.S.S.\ telescope was commissioned in June 2002. At the
time of writing, two of the four Phase~I telescopes are used for routine 
data-taking and a central trigger is being commissioned. The full system 
with four telescopes is expected to be complete and operational in early 2004.

Since June 2002, H.E.S.S.\ has recorded data both in single
telescope mode and in stereoscopic observation mode. Clear signals
are observed for the Crab nebula \cite{analysis1} and the active galactic nucleus 
PKS~2155--304 \cite{analysis2}. The latter observation confirms the
earlier detection by the Durham experiment \cite{durham}. The 
analysis of other sources is in progress.

\section{Acknowledgements}

This work was supported by the Bundesministerium f\"ur Bildung 
und Forschung under the contract number 05 CH2KHA/1.

\end{document}